\documentclass[aps,prd,twocolumn,superscriptaddress,tightenlines,nofootinbib]{revtex4-1}
\usepackage[T1]{fontenc}
\usepackage{microtype}
\usepackage[textsize=scriptsize,backgroundcolor=red!70,linecolor=red]{todonotes}
\usepackage{booktabs}
\usepackage{amsmath}
\usepackage{amsfonts}
\usepackage{amssymb}
\usepackage{hyperref}
\usepackage[all]{hypcap}
\usepackage{paralist}
\usepackage{multirow}
\usepackage{graphicx}
\usepackage{dcolumn}
\usepackage{bm}
\usepackage{epsf}

\usepackage{graphicx}
\usepackage{subfigure}
\usepackage{dcolumn}	
\usepackage{epsfig}
\usepackage{epstopdf}
\usepackage{setspace}
\usepackage{slashed}
\usepackage{comment}
\usepackage{enumitem}
\usepackage[normalem]{ulem}
\usepackage{adjustbox}

\newcommand{\PRE}[1]{{#1}} 

\newcommand{\be}{\begin{equation}\begin{aligned}}
\newcommand{\ee}{\end{aligned}\end{equation}}
\newcommand{\beq}{\begin{equation}}
\newcommand{\eeq}{\end{equation}}
\newcommand{\beqa}{\begin{eqnarray}}
\newcommand{\eeqa}{\end{eqnarray}}
\newcommand{\nn}{\nonumber}

\newcommand{\gev}{\text{GeV}}
\newcommand{\tev}{\text{TeV}}

\renewcommand{\eqref}[1]{Eq.~(\ref{#1})}

\newcommand{\order}[1]{\mathcal{O}(#1)}

\def\be{\begin{equation}}
\def\ee{\end{equation}}
\def\bea{\begin{eqnarray}}
\def\eea{\end{eqnarray}}\def\nn{\nonumber}
\def\gsim{\ \rlap{\raise 2pt\hbox{$>$}}{\lower 2pt \hbox{$\sim$}}\ }
\def\lsim{\ \rlap{\raise 2pt\hbox{$<$}}{\lower 2pt \hbox{$\sim$}}\ }
\def\dslash{\kern-4pt \not{\hbox{\kern-2pt $\partial$}}}
\def\pslash{\not{\hbox{\kern-2pt p}}}

\def\gev{{\rm GeV }}
\def\l{{\rm L}}

\definecolor{gray}{rgb}{0.90,1,1}
\definecolor{LightCyan}{rgb}{0.88,1,1}

\def \be{\beta}

\def\beq{\begin{equation}}
\def\eeq{\end{equation}}
\def\bea{\begin{eqnarray}}
\def\eea{\end{eqnarray}}
\def\ber{\begin{eqnarray*}}
\def\eer{\end{eqnarray*}}
\def\bwt{\begin{widetext}}
\def\ewt{\end{widetext}}
\def\nn{\nonumber}
\def\roughly#1{\mathrel{\raise.3ex\hbox
{$#1$\kern-.75em\lower1ex\hbox{$\sim$}}}}
\def\lsim{\roughly<}
\def\gsim{\roughly>}

\def\order{\lower 1.8ex \hbox{\LARGE\~{}}}

\usepackage{bm}

\def \({\left(}
\def \){\right)}
\def \[{\left[}
\def \]{\right]}
\def \l|{\left|}
\def \r|{\right|}

\def \nn{\nonumber}

\def \be{\beta}

\def\gev{{\rm GeV}}

\def \({\left(}
\def \){\right)}
\def \[{\left[}
\def \]{\right]}
\def \l|{\left|}
\def \r|{\right|}

\begin{document}

\preprint{UMISS-HEP-2010-02}

\title{\PRE{\vspace*{0.1 in}}
$B \to K \nu\bar\nu$, MiniBooNE
and muon $g-2$ anomalies from a dark sector
\PRE{\vspace*{.1in}}}

\author{Alakabha Datta}
\email{datta@phy.olemiss.edu}
\affiliation{Department of Physics and Astronomy, University of Mississippi, 108 Lewis Hall, Oxford, MS 38677, USA
\PRE{\vspace*{.1in}}}
\affiliation{ SLAC National Accelerator Laboratory, 2575 Sand Hill Road, Menlo Park, CA 94025, USA
\PRE{\vspace*{.1in}}}
\affiliation{Santa Cruz Institute for Particle Physics,
Santa Cruz, CA 95064, USA
\PRE{\vspace*{.1in}}}

\author{Danny Marfatia}
\email{dmarf8@hawaii.edu}
\affiliation{Department of Physics and Astronomy, University of Hawaii, Honolulu, HI 96822, USA
\PRE{\vspace*{.1in}}}

\author{Lopamudra Mukherjee}
\email{lopamudra.physics@gmail.com}
\affiliation{School of Physics, Nankai University, Tianjin 300071, China
\PRE{\vspace*{.1in}}}


\begin{abstract}
Belle~II has reported the first evidence for $B^+ \to K^+\nu\bar\nu$ with a branching ratio $2.7 \sigma$ higher than the standard model expectation. We explain this, and the MiniBooNE and muon anomalous magnetic moment anomalies in a model with a dark scalar that couples to a slightly heavier sterile Dirac neutrino and that communicates with the visible sector via a Higgs portal. 
We make predictions for rare kaon and other $B$ meson decays.


\end{abstract}


\maketitle




\textbf{Introduction.} Anomalies in the charged and neutral current $B$ decays have been an active area of research for almost a decade. 
The recently updated measured values of $\smash{R_{K^{(\ast)}} = \mathcal{B}(B\to K^{(\ast )} \mu^+\mu^-)/ \mathcal{B}(B\to K^{(\ast )} e^+ e^- )}$ are
 now fully consistent with the standard model (SM) expectation~\cite{LHCb:2022qnv}, and have dampened interest in neutral current (NC) $B$ anomalies 
(although the individual branching fractions remain discrepant~\cite{Capdevila:2023yhq}).
However, a first measurement by Belle~II of the branching ratio $\mathcal{B}(B^+\to K^+\nu\bar\nu) = (2.3\pm 0.7)\times 10^{-5}$~\cite{Belle-II:2023esi}, is $2.7\sigma$ higher than the SM expectation
$\mathcal{B}(B^+\to K^+ \nu\bar\nu)_{\rm SM}=(5.58\pm 0.38)\times 10^{-6}$~\cite{Parrott:2022zte}, and
has revived interest in NC $B$ decays~\cite{Bause:2023mfe,Allwicher:2023xyz,Athron:2023hmz,Felkl:2023ayn, he2023revisiting,chen2023flavor}; an earlier upper limit by Belle~II~\cite{Belle-II:2021rof} also led to a flurry of theoretical activity~\cite{Browder:2021hbl,He:2021yoz,Felkl:2021uxi,He:2022ljo,Ovchynnikov:2023von,Asadi:2023ucx}.
Unlike the dilepton modes in  $R_{K^{(\ast)}}$, 
the  contamination from $ c \bar{c}$ states in $B \to K^{(*)} \nu \bar{\nu}$ can be neglected. Hence, if confirmed, the Belle~II result could be a clear sign of new physics.
Note that any theoretical interpretation must be compatible with constraints on the other $B\to K^{*}\nu\bar\nu$ decays in Table~\ref{tab:constraints}.

Decays of $B$ mesons that involve invisible final states are excellent probes of new invisible or long lived
states like a massive sterile neutrino, $\nu_D$. These states may be produced via mixing with the active neutrinos or
through new mediators such as new vector bosons or leptoquarks that couple them to SM particles. 
We consider a mechanism in which $\nu_D$ communicates with the SM sector through a light scalar mediator $S$~\cite{Datta:2020auq} which couples to the SM sector through an extended Higgs portal~\cite{Batell:2016ove,Datta:2019bzu, Liu:2020qgx}. The scalar couples to active neutrinos via four-neutrino mixing
and  a contribution to  $B^+ \to K^+ \nu \bar\nu$ is generated by the two body decays, $B^+ \to K^+ S $ and
$ S \to \nu \bar{\nu}$. 

Our framework also permits an understanding of the excess in electron like events in the MiniBoone experiment~\cite{Aguilar-Arevalo:2018gpe} in terms of
a light neutrino upscattering into $\nu_D$ which subsequently decays to $\nu S(\to e^+e^-,\gamma\gamma)$.  Note that similar
 upscattering into $\nu_D$ via a dark vector boson $Z^\prime$~\cite{Bertuzzo:2018itn,Ballett:2018ynz} 
   is excluded~\cite{Arguelles:2018mtc} by data from the CHARM-II~\cite{Vilain:1994qy} and MINERvA~\cite{Valencia:2019mkf} experiments if the $Z^\prime$ is lighter than the sterile neutrino as in Ref.~\cite{Bertuzzo:2018itn}.  As shown in Ref.~\cite{Datta:2020auq}, with a light scalar mediator, the solution to the MiniBooNE anomaly
is consistent with the CHARM-II and MINERvA 
data even with $S$ lighter than the sterile neutrino.
It is noteworthy that this explanation is untested by 
the template analysis of MicroBooNE~\cite{Abdullahi:2023ejc}.

Another long-standing anomaly is that of the
anomalous magnetic moment of the muon, $a_\mu \equiv (g-2)_{\mu}/2$. 
The SM prediction~\cite{Aoyama:2020ynm} is more than $5\sigma$ smaller than the updated world average following the latest experimental measurement~\cite{Muong-2:2023cdq}:
\begin{equation}
\Delta a_\mu = a_{\mu}^{\text{exp}} - a_{\mu}^{\text{SM}} = (2.49 \pm 0.48) \times 10^{-9} \,.
\label{eq:g-2}
\end{equation} 
The anomaly can be resolved by including a higher dimensional coupling to two photons~\cite{Datta:2020auq,Datta:2019bzu}. This coupling can also help explain the MiniBooNE anomaly because it enables the scalar to decay to photon pairs which can be misidentified as electron events at MiniBooNE.

\textbf{Model.}
\label{sec:model}
The scalar Lagrangian is given by
\begin{align}
\mathcal{L}_{S} & \supset  \frac{1}{2}(\partial_\mu S)^2 \! - \! \frac{1}{2}m_{S}^2 S^2 \! - \eta_d\sum_{f=d,\ell}  \! \frac{m_f}{v} \bar{f}f  S \nn \\
& - \sum_{f = u,c,t} \! \eta_{f} \frac{ m_{f} }{v}\bar{f}f S 
- g_D S \bar {\nu}_D  \nu_D - \frac{1}{4} \kappa S F_{\mu \nu} F^{\mu \nu}\,,
\label{eq:L_S}
\end{align}
where $v \simeq 246~\gev$, is the Higgs vacuum expectation value, $d$ and $\ell$ correspond to down-type quarks and leptons with a universal coupling $\eta_d$ scaled by the respective SM Yukawa. The structure of the Lagrangian can arise from the mixing of singlet scalar with the neutral components of a two Higgs doublet model~\cite{Datta:2019bzu, Liu:2020qgx, Batell:2016ove}. We will however adopt an effective interaction in the spirit of Ref.~\cite{Datta:2020auq} and take the couplings $\eta_f$ of the scalar to the up-type quarks to not be flavor universal.
The parameter $\kappa$, of inverse mass dimension, induces an $S\gamma\gamma$ coupling which contributes to $(g-2)_\mu$ via the one-loop Barr-Zee diagram~\cite{Barr:1990vd}. (Without this higher dimensional operator, the contributions to $(g-2)_\mu$ are via the vertex correction of the $\gamma \mu^+ \mu^-$ vertex and self-energy diagrams which are insufficient to address the anomaly~\cite{Datta:2019bzu}.)

The light active neutrinos mix with the heavy sterile neutrino and induce a coupling of the dark scalar to the light neutrinos. The four flavor eigenstates $\nu_\alpha$ are related to the mass eigenstates $\nu_i$ by
\begin{align}
\nu_{\alpha(L,R)} = \sum_{i=1}^{4}U_{\alpha i} \nu_{i(L,R)}  ~, \quad \alpha=e,\mu,\tau,D\,,
\label{Eq:mixing}
\end{align}   
where $L,R$ indicate the handedness of the neutrino, and $U$ is a $4\times4$ orthogonal matrix common to $\nu_L$ and $\nu_R$. We take $U_{e4} = U_{\tau4} = 0$ which gives $1-|U_{D4}|^2=|U_{\mu4}|^2$ by unitarity. 
We require $\nu_4$ to be a Dirac neutrino so that its 
nonrelativistic decays, $\nu_4 \to \nu S$, are not isotropic~\cite{Balantekin:2018ukw}, as required by MiniBooNE data.
Note that the sterile neutrino will have a much shorter lifetime $\sim |U_{\mu4}|^{-2}$ than the scalar $\sim |U_{\mu4}|^{-4}$.

The coupling of the light scalar to up-type quarks and light neutrinos yields several flavor changing neutral current transitions via the penguin loop. The rare hadronic decays in Table~\ref{tab:constraints} provide constraints on the model.

\begin{table}[tb]
    \centering
    \renewcommand{\arraystretch}{1.7}
    \resizebox{\columnwidth}{!}{
    \begin{tabular}{|c|c|c|}
    \hline 
    Observable & SM expectation & Measurement or constraint\\
    \hline 
    $\mathcal{B}(B^+ \to K^+ \nu \bar\nu)$ & $(5.58\pm 0.38)\times 10^{-6}$ \cite{Parrott:2022zte}& $(2.3\pm 0.7)\times 10^{-5}$ ~\cite{Belle-II:2023esi}\\
    $\mathcal{B}(B^0 \to K^{*0} \nu \bar\nu)$ & $(9.2\pm 1.0)\times 10^{-6}$ \cite{Buras:2014fpa}& $< 1.8\times 10^{-5}$ ~\cite{Belle:2017oht}\\
    $\mathcal{B}(B^+ \to K^{*+}  \nu \bar\nu)$ & $\mathcal{B}(B^0 \to K^{*0} \nu \bar\nu)\frac{\tau_{B^+}}{\tau_{B^0}}$ \cite{Buras:2014fpa}& $< 4\times 10^{-5}$ ~\cite{Belle:2013tnz}\\
    $\mathcal{B}(B^0 \to K^{*0} e^+ e^-)_{0.03-1\rm{~GeV}}$ & $(2.43^{+0.66}_{-0.47})\times10^{-7}$ \cite{Jager:2012uw} & $(3.1^{+0.9+0.2}_{-0.8-0.3}\pm 0.2)\times10^{-7}$ \cite{LHCb:2013pra}\\
    $\mathcal{B}(B_s \to \gamma \gamma)$ & $5 \times 10^{-7}$ \cite{Reina:1997my}& $<3.1\times 10^{-6}$ \cite{Belle:2014sac}\\
    $\mathcal{B}(B_s \to \mu^+ \mu^-)$ & $(3.57\pm 0.17) \times 10^{-9}$ \cite{Beneke:2017vpq} & $(3.52^{+0.32}_{-0.31})\times 10^{-9}$ \cite{Neshatpour:2022pvg} \\
    $\mathcal{B}(K_{L} \to \pi^0 \nu \bar\nu)$ & $(3.4 \pm 0.6)\times 10^{-11}$ \cite{Buras:2015qea} & $< 4.9 \times 10^{-9}$ \cite{KOTO:2020prk}\\ 
    $\mathcal{B}(K_{L} \to \pi^0 e^+ e^-)$ & $(3.2^{+1.2}_{-0.8}) \times 10^{-11}$ \cite{Buchalla:2003sj} & $< 2.8 \times 10^{-10}$ \cite{AlaviHarati:2003mr}\\ 
    $\mathcal{B}(K_{L} \to \pi^0 \gamma \gamma)$ & - & $(1.273 \pm 0.033)\times 10^{-6}$ \cite{pdg2022}\\
    $\mathcal{B}(K_{S} \to \pi^0 \gamma \gamma)$ & - & $(4.9\pm 1.8)\times 10^{-8}$ \cite{pdg2022}\\
    $\mathcal{B}(K^+ \to \pi^+ \gamma \gamma)$ & - & $(1.01\pm 0.06)\times 10^{-6}$ \cite{pdg2022}\\
    $\mathcal{B}(K^\pm \to \mu^\pm \nu_\mu e^+ e^-)_{m_{e^+e^-}\geq 140~\rm{MeV}}$ & - & $(7.81 \pm 0.23)\times 10^{-8}$ \cite{Khoriauli:2017pzm}\\
    $\Delta M_{B_s}$ & $(18.4^{+0.7}_{-1.2})~\rm{ps}^{-1}$ \cite{DiLuzio:2019jyq} & $(17.765 \pm 0.006) ~\rm{ps}^{-1}$ \cite{pdg2022}\\
    $\Delta M_{K}$ & $(47\pm 18) \times 10^{8}~\rm{s}^{-1}$ \cite{Brod:2011ty} & $(52.93\pm 0.09)\times 10^8~\rm{s}^{-1}$\cite{pdg2022}\\
    $a_\mu$ & $116 591 810(43)\times 10^{-11}$ \cite{Aoyama:2020ynm}& $116 592 059(22)\times 10^{-11}$ \cite{Muong-2:2023cdq}\\
    \hline 
    \end{tabular}}
    \caption{Experimental measurements and constraints used in the analysis. The upper limits are at 90\% C.L.}
    \label{tab:constraints}
\end{table}

\textbf{Signals and constraints.} We consider a dark scalar with mass in the range $10 \lsim m_S/{\rm MeV} \lsim 150$. Since $m_S < 2m_\mu$, $S$ can only decay to photons, electrons and neutrinos, for which the decay widths are provided in the Supplemental Material. Below, we discuss the phenomenological implications for several observables of interest and the constraints imposed on the model.

\textit{S decay length.} We require the decay length of the dark scalar to be shorter than 0.1~mm to evade bounds from beam dump and other experiments that probe long-lived particles.

\textit{B and K meson decays.} The Lagrangian in \eqref{eq:L_S} induces two-body meson decays such as \mbox{$B \to K^{(*)} S$} and $K \to \pi S$ at one loop. The flavor changing neutral current (FCNC) Lagrangian for the $b\to s$ and $s\to d$ transitions is given by
\beq
\mathcal{L}_{\rm FCNC} = g_{bs} \bar{s}P_R b S + g_{sd}\bar{d}P_R s S\,,
\label{eq:L-FCNC}
\eeq 
where the effective couplings $g_{bs}$, $g_{sd}$ take the form,
\beq
g_{bs} \approx \frac{3\sqrt{2}G_F}{16\pi^2}\frac{m_t^2 m_b }{v}\eta_t V_{tb}V_{ts}^*\,,
\eeq 
and
\beq
g_{sd} \approx \frac{3\sqrt{2}G_F}{16\pi^2}\frac{m_t^2 m_s }{v}V_{ts}V_{td}^*\(\eta_t + \eta_c \frac{m_c^2}{m_t^2}\frac{V_{cs}V_{cd}^*}{V_{ts}V_{td}^*}\)\,.
\eeq
The rates for the two body FCNC processes can be found in the Supplemental Material. The relevant $B$ meson form factors are taken from Refs.~\cite{Ball:2004ye, Straub:2015ica} while the kaon form factor $|f_0^{K}(m_S^2)|^2 $ is approximately unity~\cite{Marciano:1996wy}. It is important to note that $g_{bs}$ is essentially determined by $\eta_t$ since the contribution proportional to $\eta_c$ is both helicity and charm-mass suppressed. On the other hand, for $g_{sd}$, the contribution from $\eta_c$ is only helicity suppressed and can be sizable for $\eta_c \gg \eta_t$. $\mathcal{L}_{\rm FCNC}$ also induces $B_s$ decays to  $\gamma \gamma,\, \mu^+ \mu^-$ and $\nu \bar \nu$. The most important constraints on the couplings from the flavor changing $B$ and $K$ transitions are as follows:
\begin{enumerate}
    \item $B$ decay width: We require $\mathcal{B}(B \to K^{(*)}S) < 10\%$ so that it does not exceed the uncertainty in the SM prediction of the $B$ meson width~\cite{Lenz:2014jha}.
    \item $B \to K \nu \bar\nu$: We require $\mathcal{B}(B^+ \to K^+ \nu \bar\nu)$ to lie within $1\sigma$ of the Belle~II measurement in the first row of Table~\ref{tab:constraints}. 
    \item $B \to K^{*} \nu \bar\nu$ : We ensure that the upper limits on the branching fractions of the $B \to K^*$ modes in Table~\ref{tab:constraints} are satisfied.
    \item $B^0 \to K^{*0} e^+ e^-$: This decay has been measured by LHCb in the low dilepton mass region of {\mbox 30-1000 MeV/c$^2$}~\cite{LHCb:2013pra} which overlaps  the mass range of the dark scalar. We therefore require the branching ratio to lie within $1\sigma$ of the measured value.
    \item $B_s$ decays: We require the scalar contribution to $B_s \to \gamma \gamma$ to remain below the upper limit placed by Belle~\cite{Belle:2014sac} and the contribution to $B_s \to \mu^+ \mu^-$ to remain within the 1$\sigma$ uncertainty of the measurement in Table~\ref{tab:constraints}.
    We take the decay constant $f_{B_s} = 230.3(1.3)\rm{MeV}$ from Ref.~\cite{FlavourLatticeAveragingGroupFLAG:2021npn}.
    An interesting signature is $B_s \to \rm{invisible}$ which is currently not constrained by experiment but can be probed at Belle~II. We make predictions for its branching fraction for some benchmark points. 
    \item $K^+ \to \pi^+ \nu \bar\nu$: The NA62 experiment at CERN has set stringent limits on $\mathcal{B}(K^+ \to \pi^+ X)$ as a function of the $X$ mass and lifetime for invisible $X$ decays except in the range $110<m_X/\rm{MeV}<160$~\cite{NA62:2021zjw, NA62:2020pwi}. However, since $S$ has a very short lifetime, $\mathcal{O}(0.1)$~ps, these limits do not apply. 
    \item $K_{L} \to \pi^0 \nu \bar\nu$: We require that the most recent upper limit on the branching fraction from the KOTO experiment at J-PARK be satisfied; see Table~\ref{tab:constraints}.
    \item $K \to \pi \gamma \gamma$: For these decay modes, we require the dark scalar contribution to be smaller than the measured central values in Table~\ref{tab:constraints}.
    \item $K^\pm \to \mu^\pm \nu_\mu e^+ e^-$: The NA48/2 Collaboration has measured $\mathcal{B}(K^\pm \to \mu^\pm \nu_\mu e^+ e^-) = (7.81 \pm 0.23) \times 10^{-8}$ \cite{Khoriauli:2017pzm} in the kinematic region $m_{e^+e^-} \geq 140$~MeV. In our model this decay proceeds through $K \to \mu \nu S (\to e^+ e^-)$ where the dark scalar is radiated off the muon leg. The constraint applies for \mbox{$m_S > 140$~MeV}.
\end{enumerate} 

\textit{B and K meson mixing.} The Lagrangian in \eqref{eq:L-FCNC} also induces meson mixing. The measurement of the mass difference $\Delta M_{B_s}$ is consistent with the SM prediction. Hence we require the additional contribution to not exceed the uncertainty in the SM expectation. For kaon mixing, the SM prediction suffers from large uncertainties. The long distance contribution is poorly estimated, so we only include the short distance contribution to $\Delta M_K$ in Table~\ref{tab:constraints}. The measured value, however, is quite precise and we do not allow the new contribution to $\Delta M_K$ to exceed the $1\sigma$ uncertainty in the measurement.

\textit{$(g-2)_\ell$.} The dominant contribution to the anomalous magnetic moments of the muon and electron comes from the log enhanced term of the Barr-Zee diagram (see Supplemental Material) which depends on the cutoff scale $\Lambda$. We fix $\Lambda=2~\tev$ and require consistency with Eq.~(\ref{eq:g-2}) within $1\sigma$. Because $\Delta a_e/\Delta a_\mu = m_e^2/m_\mu^2$, we find $\Delta a_e \sim \rm{few}\times 10^{-14}$ which is much smaller than the magnitude of the inferred value, ${\cal{O}} (10^{-13}-10^{-12})$~\cite{Hanneke:2008tm,Parker:2018vye,Aoyama:2019ryr}.

\textit{MiniBooNE.} The scattering of a muon neutrino off a nucleus may take place via the dark scalar exchange as shown in Fig.~\ref{fig:Miniboone}. Being heavy, the sterile neutrino produced promptly decays to a light neutrino and the dark scalar, whose subsequent decay to $e^+e^-$ and $\gamma\gamma$ may mimic the signal observed in the MiniBooNE experiment. 
Since MiniBooNE is a Cherenkov detector that cannot distinguish between electrons and photons, two photons or two electrons with a small opening angle can be misidentified as a single electron.
The details of the coherent scattering cross sections mediated by the dark scalar can be found in the Supplemental material. 
\begin{figure}[t]
    \centering
    \includegraphics[scale=0.32]{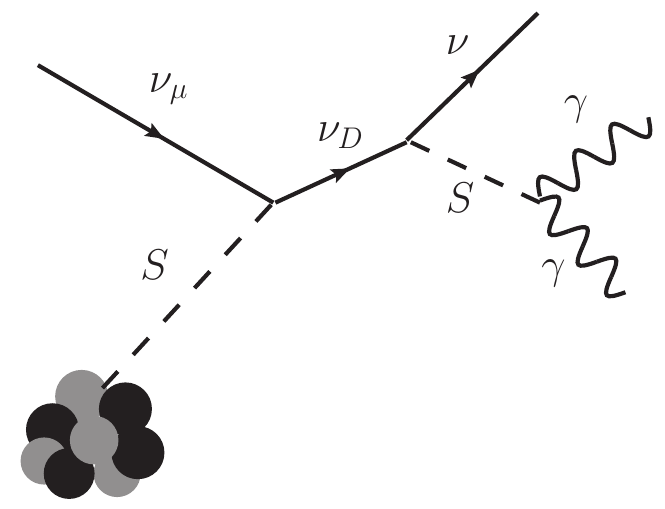}~~~~~
    \includegraphics[scale=0.32]{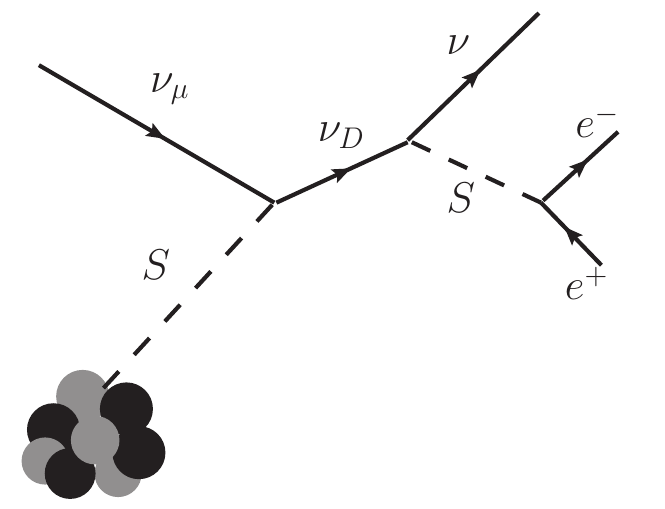}
    \caption{Dark scalar mediated neutrino-nucleus scattering produces the MiniBooNE signal.}
    \label{fig:Miniboone}
\end{figure}

It has been previously shown in Ref.~\cite{Datta:2020auq} that the dark scalar model is able to explain the excess observed in MinibooNE data while being consistent with the observations of the CHARM II experiment for $m_{\nu 4}$ between 400~MeV and 500~MeV. The sterile neutrino has to be heavier than 400~MeV to ensure that less than 70\% of the excess events are in the
    forward most bin $(0.8 < \cos \theta < 1)$ of the angular distribution of electron-like events at
    MiniBooNE \cite{Arguelles:2018mtc}. We compute the coherent and incoherent scattering cross sections at MiniBooNE and CHARM-II and recast the results of Refs.~\cite{Bertuzzo:2018itn, Arguelles:2018mtc} for a dark $Z^\prime$ kinetically mixed with the electromagnetic field, to our case. The following constraints are imposed to implement the mapping between the dark $Z^\prime$ model and our model:
    \begin{enumerate}
        \item In terms of the total scattering cross sections $\sigma_S$ and $\sigma_{Z^\prime}$ 
        for the scalar and $Z^\prime$ mediators, respectively, we define,
        \beq 
        \small
       \mathcal{R} =\frac{\int \Phi \frac{d\sigma_S}{dT} dT dE_{\nu_\mu} \times (\mathcal{B}(S \to e^+e^-)+\mathcal{B}(S\to \gamma \gamma))}{\int \Phi \frac{d\sigma_{Z^\prime}}{dT} dT dE_{\nu_\mu}\times \mathcal{B}(Z^\prime \to e^+e^-)}\,,
        \eeq 
        with the denominator evaluated at the benchmark point $m_{Z^\prime} = 30~\rm{MeV},~\alpha_{Z^\prime} = 0.25,~\alpha \epsilon^2 =2\times10^{-10}$ of Ref.~\cite{Bertuzzo:2018itn} to explain the MiniBooNE anomaly. The $\nu_\mu$ flux at the Booster Neutrino Beam in the neutrino run~\cite{Aguilar-Arevalo:2018gpe} is denoted by $\Phi$. To reproduce the MiniBooNE signal we require $0.95 \leq \mathcal{R} \leq 1.05$.
        \item The $Z^\prime$ model of Ref.~\cite{Bertuzzo:2018itn}, however, is excluded by the CHARM-II constraint in Ref.~\cite{Arguelles:2018mtc}. To satisfy this constraint we require 
        \beq {{\sigma_S \times (\mathcal{B}(S \to e^+e^-)+\mathcal{B}(S\to \gamma \gamma))}\over {\sigma_{Z^\prime} \times\mathcal{B}(Z^\prime \to e^+e^-)}} <1
        \eeq 
        for $E_{\nu_\mu} = 20$~GeV, where the denominator is evaluated for the parameter values in Fig.~3 of Ref.~\cite{Arguelles:2018mtc} with $|U_{\mu 4}| = 10^{-4}$.
    \end{enumerate}

\begin{figure*}[t]
    \centering
    \includegraphics[width=0.32\linewidth]{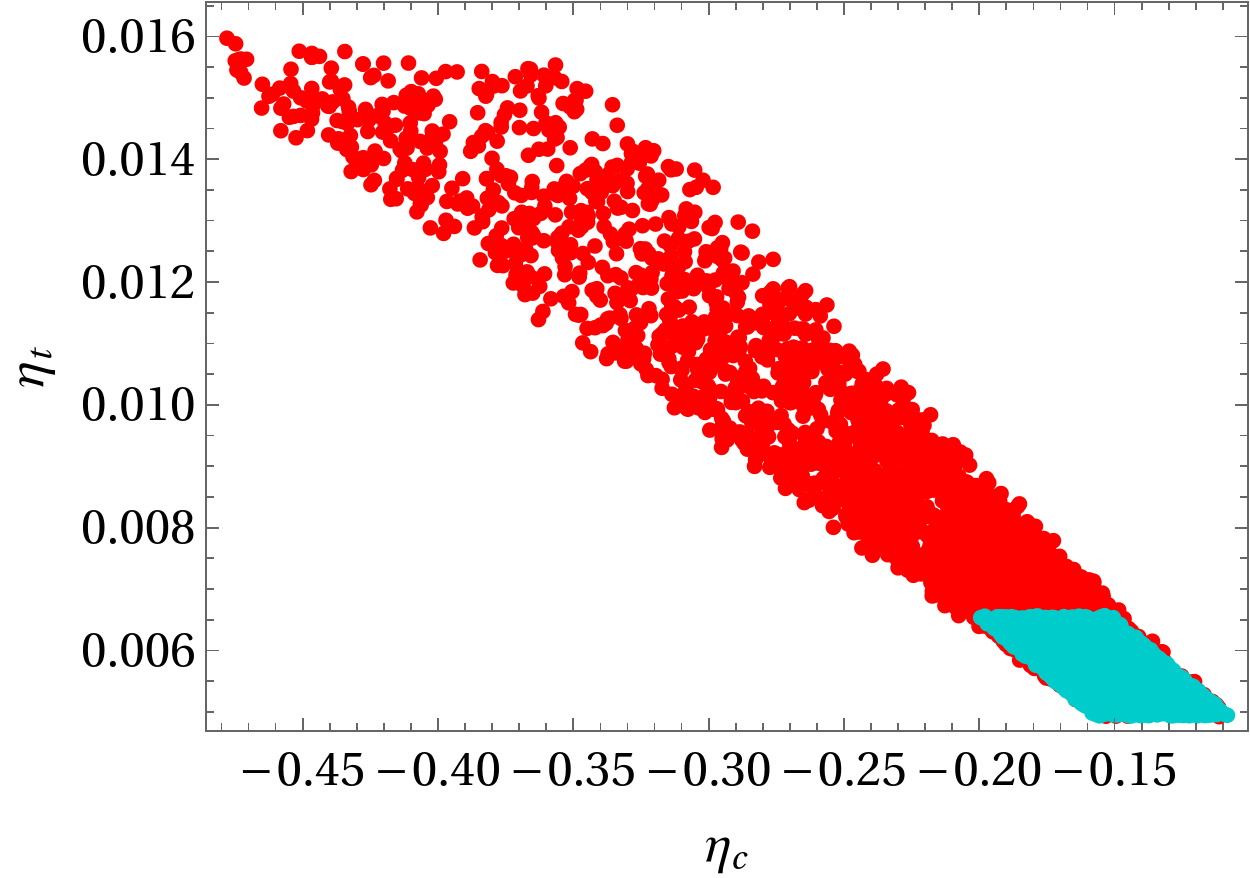}
    \includegraphics[width=0.307\linewidth]{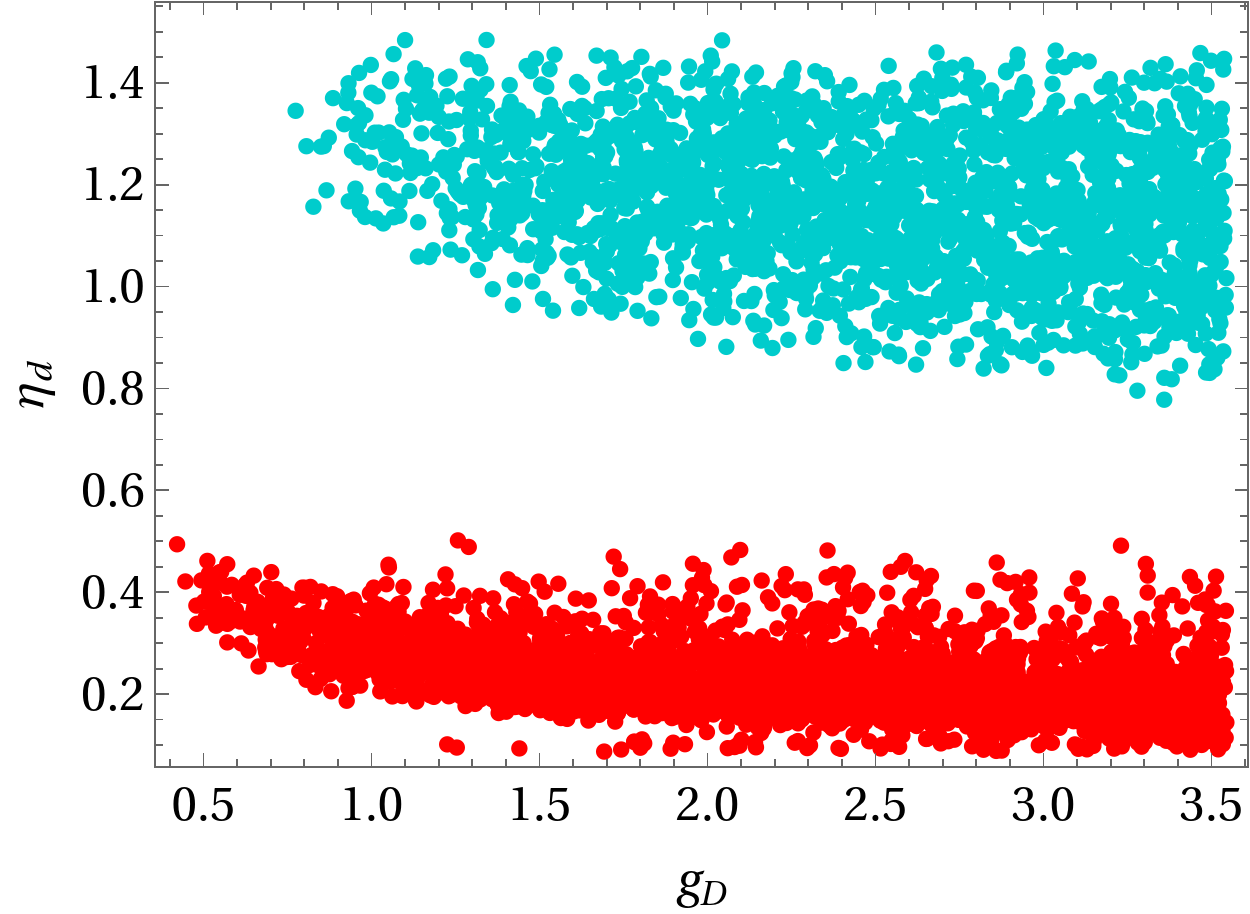}
    \includegraphics[width=0.32\linewidth]{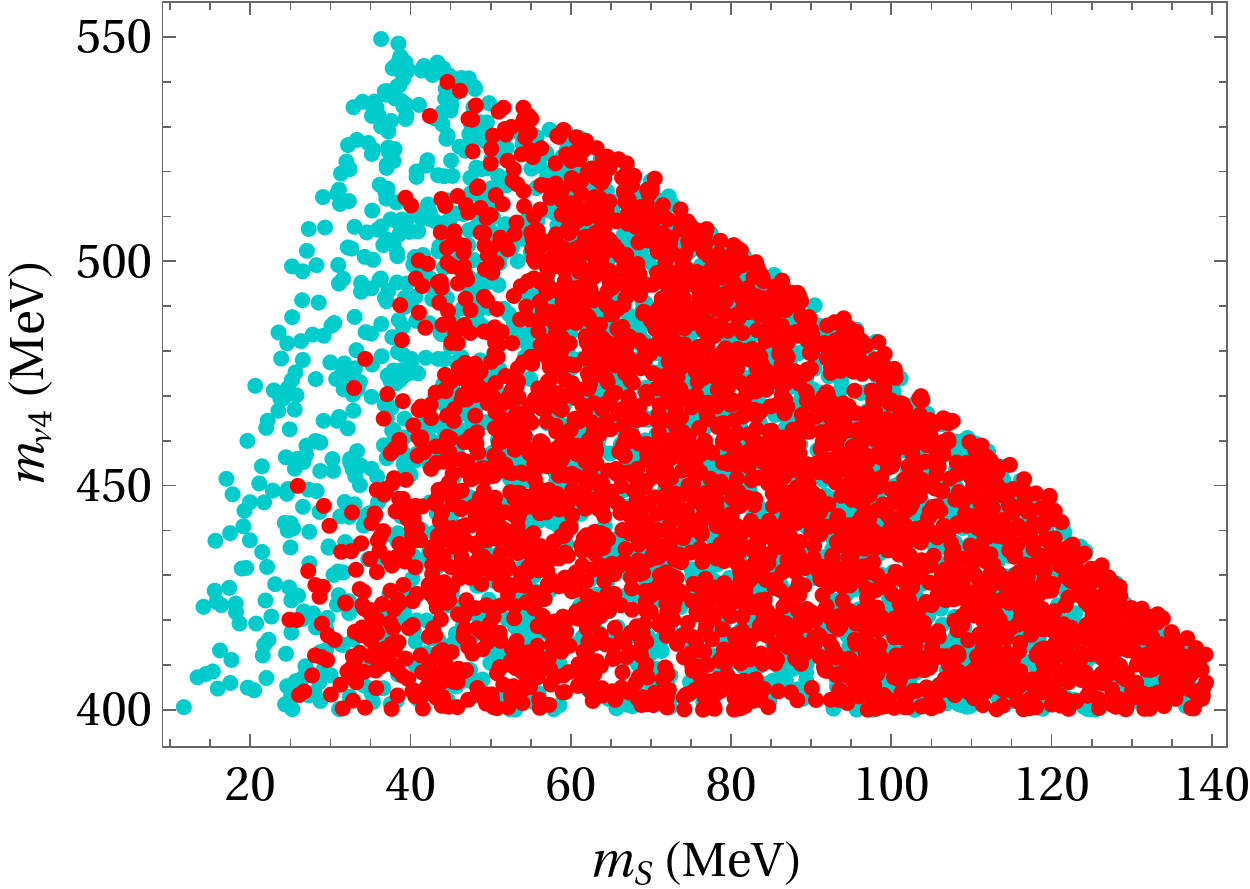}\\
    \includegraphics[width=0.32\linewidth]{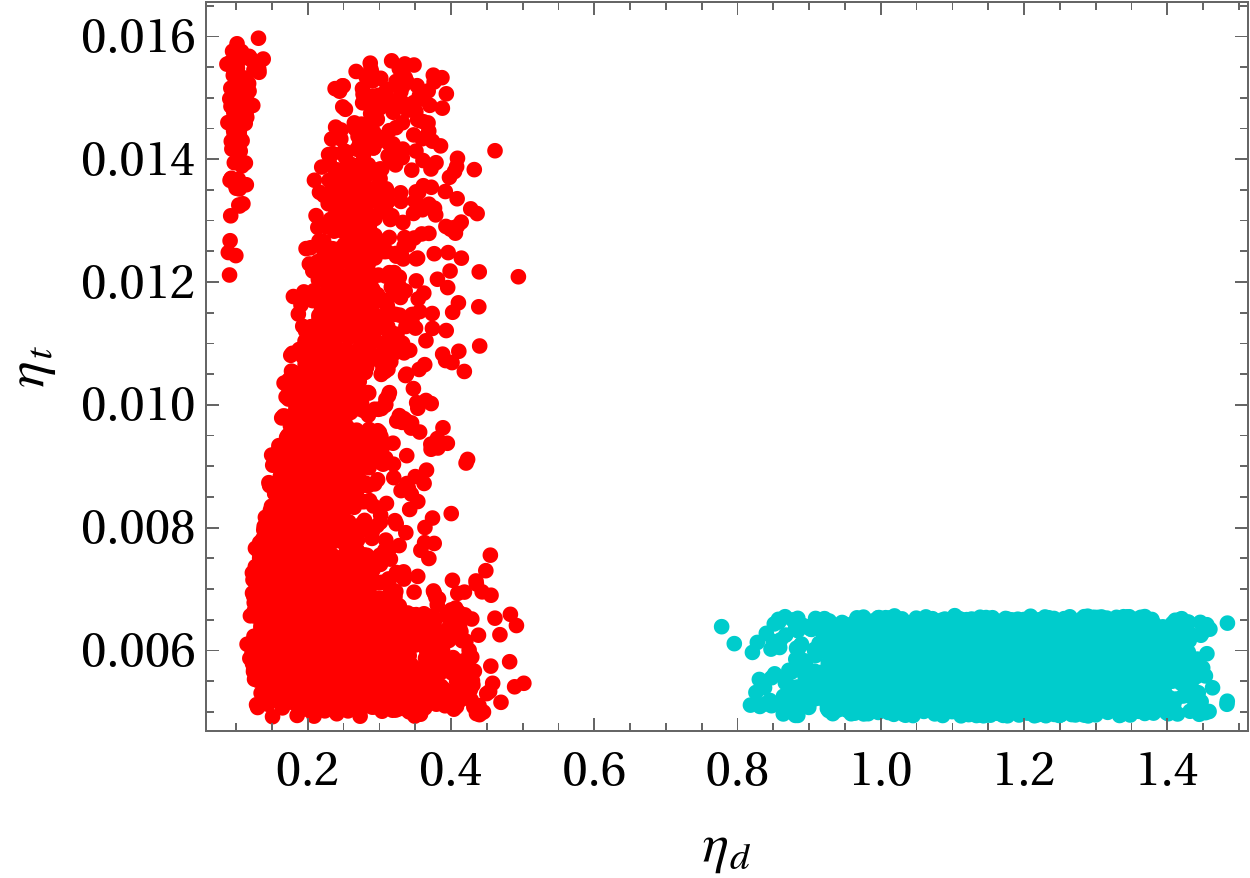}
    \includegraphics[width=0.305\linewidth]{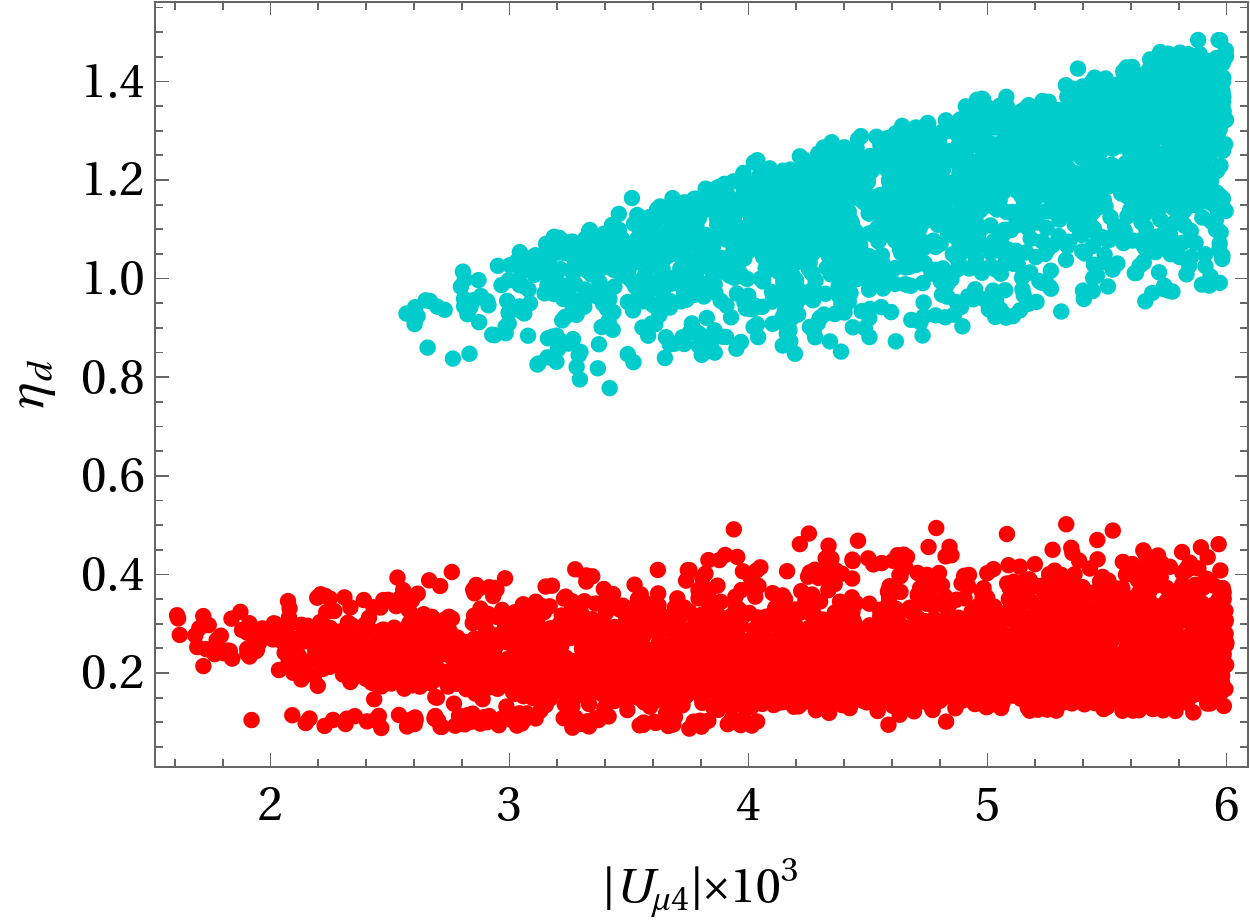}
    \includegraphics[width=0.31\linewidth]{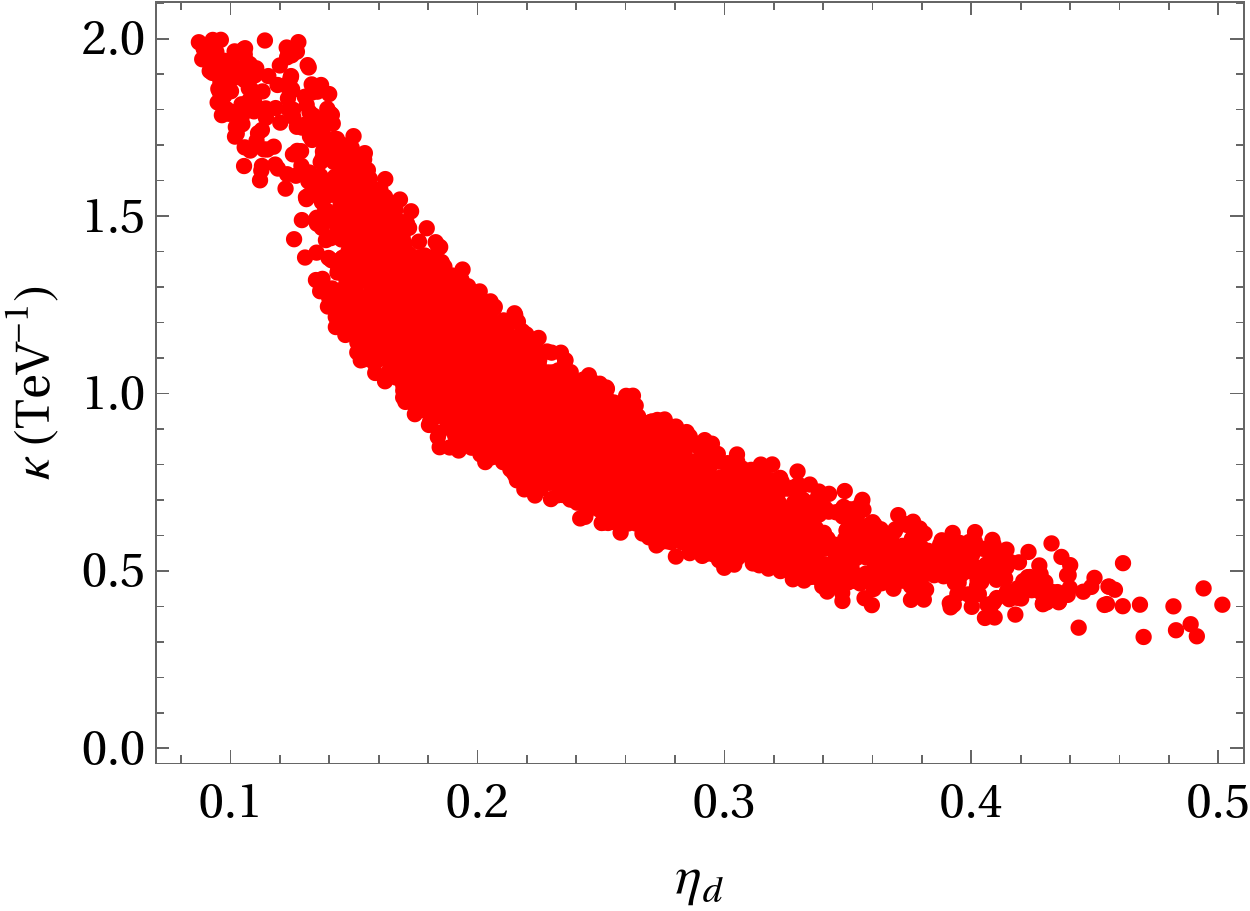}
    \caption{Allowed regions depicted by red points for $\kappa \neq 0$ and cyan for $\kappa =0$.}
    \label{fig:param-space}
\end{figure*}

\textit{Heavy neutral lepton searches.} Several experiments including PS191 \cite{Bernardi:1987ek}, NuTeV \cite{NuTeV:1999kej}, BEBC \cite{WA66:1985mfx}, FMMF \cite{FMMF:1994yvb}, CHARM~II~\cite{CHARMII:1994jjr}, T2K~\cite{T2K:2019jwa},
NA62~\cite{NA62:2021bji} and 
MicroBooNE~\cite{MicroBooNE:2023icy} have placed limits on heavy neutral leptons with sufficiently long lifetimes that can reach the detector before decaying into SM particles.  However, the limits on $U_{\mu4}$ from the nonobservation of the decay signal do not apply to our model because $\nu_4$ has a lifetime of ${\cal{O}}(0.1)$~ps and decays promptly. 
The only relevant bound for $m_{\nu4}$ between 400~MeV and 700~MeV arises from lepton universality: $|U_{\mu4}| \lsim  0.07$ at the 
99\%~C.L.~\cite{deGouvea:2015euy}.

\textbf{Analysis and results.} We analyze two cases: \mbox{$\kappa \neq 0$} and $\kappa = 0$. We find that $\eta_t \gtrsim 0.005$ is needed to explain the measured branching fraction of $B^+ \to K^+\nu\bar\nu$. However, this leads to a correction to the $K_L \to \pi^0 \nu \bar\nu$ branching fraction that violates the upper limit set by the KOTO experiment. To lower the new contribution to $K_L \to \pi^0 \nu \bar\nu$, a cancellation between the $\eta_t$ and $\eta_c$ dependent terms in $g_{sd}$ is required. This is achieved for $\eta_c \approx -27 \eta_t$. The large $\eta_c$ does not, however, affect $g_{bs}$ significantly due to the charm mass suppression. If nonzero, the coupling $\kappa$ is primarily constrained by MiniBooNE and $g-2$ data. Even though the FCNC transitions to invisible final states depend upon the sterile neutrino parameters $g_D$ and $U_{\mu 4}$, these are mainly constrained by the $\nu_\mu$-nucleus scattering cross section. For \mbox{$\kappa \neq 0$} and $\kappa = 0$, we set $\eta_u = 0$ for simplicity and scan the other
parameters 
in the following ranges:
\begin{align*}
    &\eta_d \in \[0,1\]\,,~\eta_t \in \[0,0.02\]\,,~\eta_c \in \[-0.5,0\]\,,\\
    & g_D \in \[0, \sqrt{4\pi}\]\,,~\kappa \in \[0,2\]~\rm{TeV}^{-1}\,,~|U_{\mu 4}| \in \[0, 0.006\]\,,\\
    & m_S \in \[10, 150\]~{\rm{MeV}}\,,~m_{\nu 4} \in \[400,700\]~\rm{MeV}\,.
\end{align*}

For both cases, we choose some benchmark points (BPs) that have interesting consequences for rare $K$ and $B$ decays.

\paragraph{$\kappa \neq 0$:} In this case, the dark scalar has a nonzero effective coupling to photons which permits an explanation of the discrepancy in the muon anomalous magnetic moment. The  MiniBooNE signal is also enhanced due to the nonzero branching fraction to $\gamma \gamma$. In Fig.~\ref{fig:param-space}, we show the allowed regions by the red points. As explained above, there is a strong correlation between $\eta_t$ and $\eta_c$. A significant correlation is also observed between $m_S$ and $m_{\nu 4}$. For certain values of $\eta_d$, the effective scalar-nucleus coupling responsible for neutrino-nucleus scattering becomes too small to produce a significant scattering cross section due to the fine-tuning between $\eta_t$ and $\eta_c$. Hence we observe gaps in the allowed regions of $\eta_d$.

We select five benchmark points, as listed in Table~\ref{tab:bp}, that simultaneously explain the $B^+ \to K^+\nu\bar\nu$, MiniBooNE and $g-2$ anomalies. For each BP we also provide predictions for some important decays in Table~\ref{tab:predictions}.

\begin{table}[t]
    \centering
    \renewcommand{\arraystretch}{1.5}
    \resizebox{\columnwidth}{!}{
    \begin{tabular}{|c|c|c|c|c|c|c|c|c|}
    \hline 
    BP & $\kappa$ (TeV$^{-1}$) & $\eta_d$ & $\eta_c$ & $\eta_t \times 10^2$ & $g_D$ & $U_{\mu 4} \times 10^3$ & $m_S$ (MeV) & $m_{\nu 4}$ (MeV)\\
    \hline
    1 & 1.22 & 0.17 & -0.13 & 0.54 & 2.18 & 4.86 & 38 & 413 \\
    2 & 0.60 & 0.30 & -0.28 & 1.04 & 0.94 & 3.64 & 93 & 413 \\
    3 & 1.21 & 0.21 & -0.24 & 0.84 & 1.26 & 5.65 & 93 & 432 \\
    4 & 1.03 & 0.20 & -0.14 & 0.58 & 3.42 & 3.25 & 44 & 514 \\
    5 & 0.54 & 0.44 & -0.34 & 1.31 & 0.52 & 4.82 & 138 & 404 \\
    \hline  
    6 & 0 & 0.88 & -0.13 & 0.54 & 3.06 & 4.73 & 24 & 401 \\
    7 & 0 & 1.40 & -0.15 & 0.63 & 1.82 & 5.87 & 122 & 429 \\
    \hline  
    \end{tabular}}
    \caption{Benchmark points for $\kappa \neq 0$ and $\kappa = 0$ (in which case the $g-2$ anomaly is unsolved).}
    \label{tab:bp}
\end{table}

\begin{table}[t]
    \centering
    \renewcommand{\arraystretch}{1.6}
    \resizebox{\columnwidth}{!}{
    \begin{tabular}{|c|c|c|c|c|c|c|c|c|c|}
    \hline 
    BP & $\mathcal{B}(S \to \gamma \gamma)$ & $\mathcal{B}(S \to \nu \bar \nu)$ & $\mathcal{B}(S \to e^+ e^-)$ & $\mathcal{B}(K_L \to \pi^0 \nu \bar\nu)$ & $\mathcal{B}(B_s \to \nu \bar\nu)$ & $\mathcal{B}(B \to K^{(*)} \gamma \gamma)$ \\
    \hline
    1 & 0.093 & 0.907 & $4.26 \times 10^{-5}$ &  $1.71 \times 10^{-9}$ & $5.13 \times 10^{-7}$ & $1.3 \times 10^{-6}$ \\
    2 & 0.717 & 0.282 & $7.06\times 10^{-4}$ & $3.61 \times 10^{-11}$ & $3.54 \times 10^{-7}$ & $3.7 \times 10^{-5}$ \\
    3 & 0.496 & 0.504 & $5.93\times 10^{-5}$ &  $9.02 \times 10^{-10}$ & $4.14 \times 10^{-7}$ & $1.7 \times 10^{-5}$ \\
    4 & 0.165 & 0.835 & $1.10\times 10^{-4}$ &  $1.73 \times 10^{-9}$ & $1.43 \times 10^{-6}$ & $2.65\times 10^{-6}$ \\
    5 & 0.829 & 0.170 & $9.72\times 10^{-4}$ &  $2.04 \times 10^{-10}$ & $1.72 \times 10^{-7}$ & $6.8 \times 10^{-5}$ \\
    \hline 
    6 & $4.58\times 10^{-6}$ & 0.999 & $7.10\times 10^{-4}$ &  $ 1.89\times 10^{-9}$ & $ 1.01\times 10^{-6}$ & $6.5 \times 10^{-11}$ \\
    7 & $3.95\times 10^{-4}$ & 0.997 & $2.14\times 10^{-3}$ &  $2.84 \times 10^{-9}$ & $ 4.86\times 10^{-7}$ & $7.6 \times 10^{-9}$ \\
    \hline 
    \end{tabular}}
    \caption{Predictions for the benchmark points in Table~\ref{tab:bp}.}
    \label{tab:predictions}
\end{table}
\vspace{0.1in}
\noindent

\begin{figure}[t]
    \centering
    \includegraphics[width=0.47\linewidth]{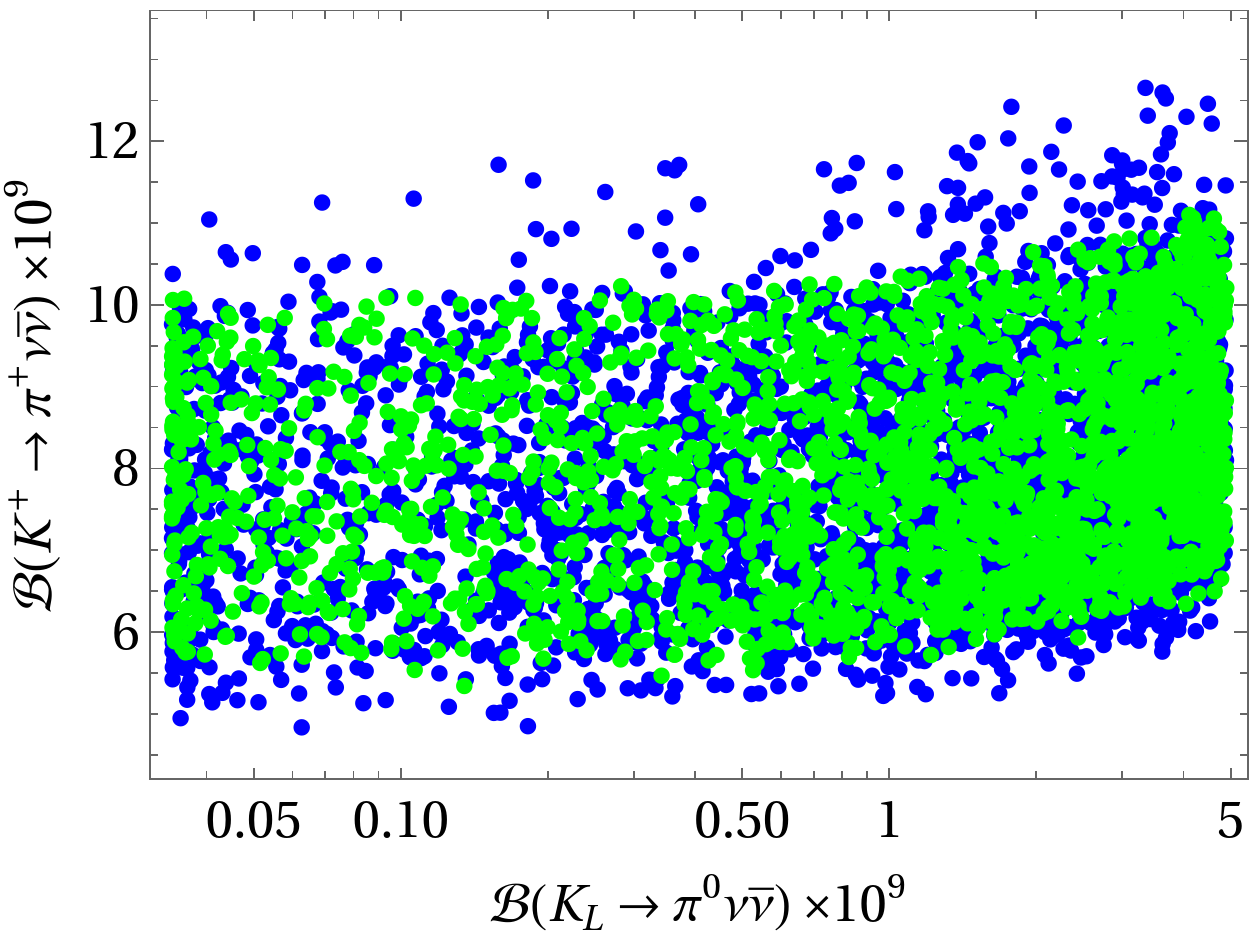}
    \includegraphics[width=0.5\linewidth]{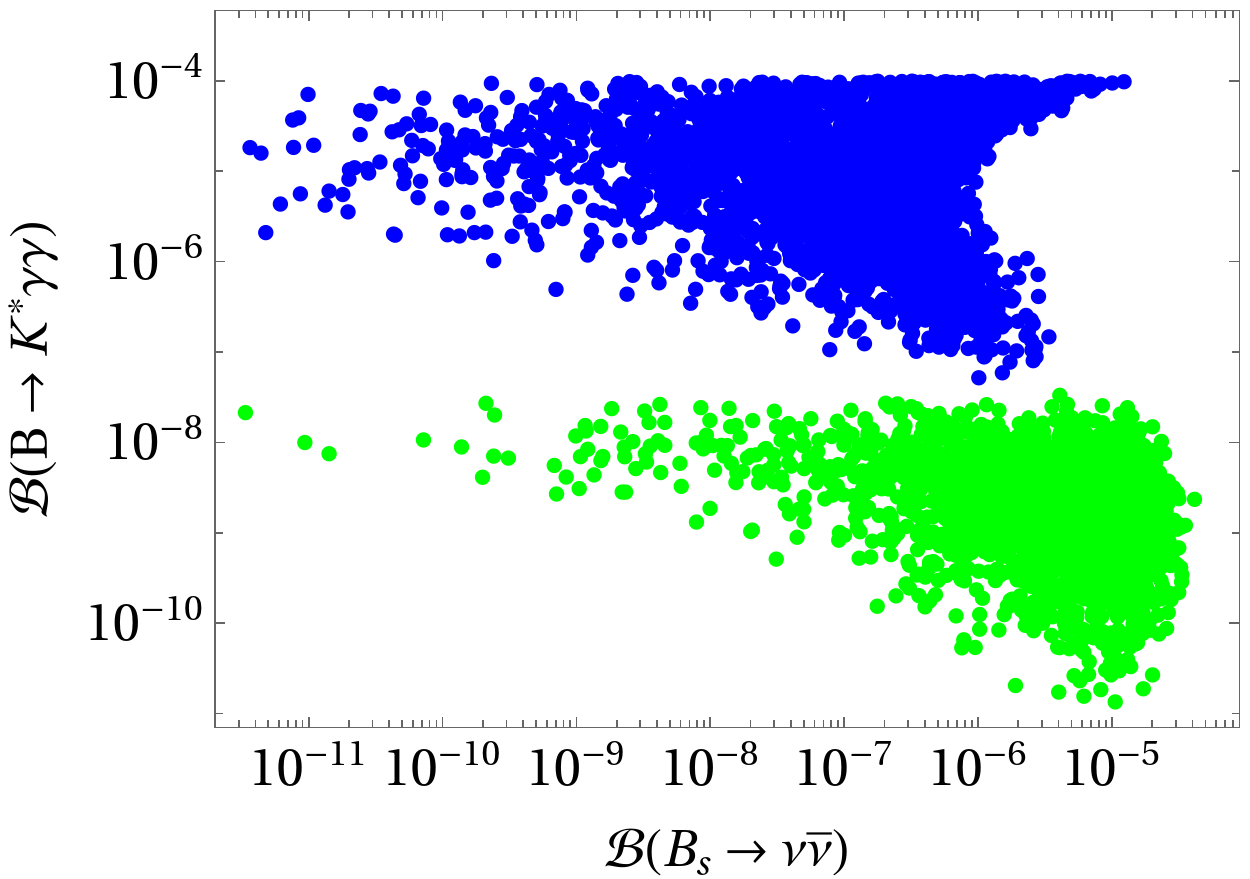}
    \caption{Branching fractions for kaon and $B$ meson decays for $\kappa\neq 0$ (blue) and $\kappa =0$ (green).}
    \label{fig:branching}
\end{figure}

\paragraph{$\kappa = 0$:} The allowed regions are shown by the cyan points in Fig.~\ref{fig:param-space}. Since the branching of $S \to \nu \bar \nu$ is enhanced in the absence of the $\gamma\gamma$ mode, $\eta_t$ is restricted to smaller values by 
$\mathcal{B}(B \to K^{(*)} \nu\bar\nu)$. The correlation of $\eta_t$ with $\eta_c$ remains the same as in the previous case. Further, the contribution to the MiniBooNE signal now only arises from the decay of the dark scalar to $e^+ e^-$. Hence, larger values of $\eta_d$ are required to enhance the overall coherent scattering cross section. Smaller values of $g_D$ and the mixing parameter $U_{\mu 4}$ are also disfavored. Two benchmark points and predictions for them are
provided in Tables~\ref{tab:bp} and~\ref{tab:predictions}, respectively.

Correlations between the branching fractions for kaon and $B$ meson decays are shown in Fig.~\ref{fig:branching}. 
The branching fraction for $K^+ \to \pi^+\nu\bar\nu$ is predicted to lie in a quite narrow range.
Note that \mbox{$K^+ \to \pi^+\nu\bar\nu$} and $B_s \to \nu\bar\nu$ are currently unconstrained
for a short-lived dark scalar and sterile neutrino.

\textbf{Summary.} Motivated by the recent Belle~II evidence for 
$B^+ \to K^+ \nu\bar\nu$ which shows a $\sim 3 \sigma$ excess relative to the SM prediction, we explored a new physics explanation of the result in terms of new light states.
The new physics contribution to this decay is interpreted as
$ B^+ \to K^+ S$, where $S$ is a light scalar in the mass range $10 \lsim m_S/{\rm MeV} \lsim 140$ which then decays to light neutrinos.
The interaction of $S$ with light neutrinos occurs via its coupling  to a heavy neutral lepton $\nu_D$ which mixes with the light neutrinos. We demonstrated that our model is consistent with other measurements and bounds and has interesting predictions for  several $B$ and $K$ decays. The excess in electron-like events observed by the MiniBooNE experiment is explained by the upscattering of $\nu_\mu$ to $\nu_D$ which subsequently decays
to $e^+ e^-$ and $\gamma \gamma$ states. Finally, our model can accommodate the recent muon $g-2$ measurement if $S$ directly couples to photons.

\vspace{0.1 in}
{\bf Acknowledgments}.  
A.D. thanks the SLAC National Accelerator Laboratory and the Santa Cruz Institute of Particle Physics
for their hospitality during the completion of this work.
A.D. is supported in part by the U.S. National Science Foundation under Grant No.~PHY-2309937 D.M. is supported in part by the U.S. Department of Energy under Grant No.~de-sc0010504.

\newpage
\bibliography{KOTO_new}

\clearpage
\newpage
\maketitle
\onecolumngrid

\begin{center}
	\textbf{$B \to K \nu\bar\nu$, MiniBooNE
and muon $g-2$ anomalies from a dark sector} \\
	\vspace{0.05in}
	{ \it \large Supplemental Material}\\
	\vspace{0.05in}
	{Alakabha Datta, Danny Marfatia and Lopamudra Mukherjee}
\end{center}

\onecolumngrid
\setcounter{equation}{0}
\setcounter{figure}{0}
\setcounter{table}{0}
\setcounter{section}{0}
\setcounter{page}{1}
\makeatletter
\renewcommand{\theequation}{S\arabic{equation}}
\renewcommand{\thefigure}{S\arabic{figure}}
\renewcommand{\thetable}{S\arabic{table}}
\newcommand\ptwiddle[1]{\mathord{\mathop{#1}\limits^{\scriptscriptstyle(\sim)}}}

\section{Decay widths}
\label{sec:rates}
We collect expressions for the relevant decay widths.

The decay width of $S$ to $e^+e^-$ is
\begin{align}
\Gamma_{S\to e^+e^-} = \frac{\eta_d^2}{8\pi} \frac{m_e^2 m_S}{v^2} \left( 1-4{m_e^2\over m_S^2} \right)^{3/2}\,,
\end{align}
while its decay width to all three light neutrinos is
\begin{align}
\Gamma_{S\to\nu\nu} = \frac{g_D^2}{8\pi}(1-|U_{D4}|^2)^2m_S\,.
\end{align}

The decay width of the dark scalar to two photons is 
\begin{align}
\Gamma_{S\to \gamma \gamma} = \frac{\alpha_{\rm EM}^2 m_S^3}{1024\pi^3} \left|\frac{4\pi}{\alpha_{\rm EM}}\kappa + \sum_{f=u,c,t,d} \frac{6\eta_f}{v} Q_f^2 A_{1/2}(r_f)\right|^2\,,
\end{align}
where $f$ denotes all spin 1/2 fermions with electric charge $Q_f$, and $r_f = 4m_f^2/m_S^2$. The loop function $A_{1/2}$ is given by~\cite{Carena:2012xa}
\begin{align}
A_{1/2}(x) = 2x^2 [ x^{-1}+(x^{-1}-1) f(x^{-1})]\,,
\end{align}
with $f(x) = \rm{arcsin}^2 \sqrt{x}$ for $m_S < 2 m_f$.

The decay width of $\nu_4$ to $S\nu$ (where $\nu$ denotes all three light neutrinos) is
\begin{equation}
\Gamma_{\nu_4 \to S \, \nu} = \frac{g_D^2}{8 \pi}|U_{D4}|^2 \left( 1-|U_{D4}|^2 \right) \left(1-{m_S^2\over m_{\nu_4}^2}  \right)^2 m_{\nu_4}\,.
\end{equation}

\section{Meson decays}
\label{sec:meson-decay}
The branching fractions for the different exclusive FCNC processes of interest are given by \cite{Datta:2019bzu}
\bea
    \mathcal{B}(B \to K S) &=& \frac{|g_{bs}|^2 f_0^2(m_S^2)\sqrt{\lambda(m_B^2,m_K^2,m_S^2)}}{64\pi m_B^3}\(\frac{m_B^2-m_K^2}{m_b-m_s}\)^2 \tau_B,\\
\mathcal{B}(B \to K^* S) &=& \frac{|g_{bs}|^2 A_0^2(m_S^2)\lambda(m_K^2,m_\pi^2,m_S^2)^{3/2}}{64\pi m_B^3(m_b+m_s)^2}\tau_B,\\
\mathcal{B}(K \to \pi S) &=& \frac{|g_{sd}|^2 f_{0,K}^2(m_S^2)\sqrt{\lambda(m_B^2,m_K^2,m_S^2)}}{64\pi m_K^3}\(\frac{m_K^2-m_\pi^2}{m_s-m_d}\)^2 \tau_K,\\
\mathcal{B}(K_L \to \pi^0 S) &=& \frac{\[\rm{Re}(g_{sd})\]^2 f_{0,K}^2(m_S^2)\sqrt{\lambda(m_{K^0}^2,m_{\pi^0}^2,m_S^2)}}{64\pi m_{K^0}^3}\(\frac{m_{K^0}^2-m_{\pi^0}^2}{m_s-m_d}\)^2 \tau_{K_L},\\
\mathcal{B}(K_S \to \pi^0 S) &=& \frac{\[\rm{Im}(g_{sd})\]^2 f_{0,K}^2(m_S^2)\sqrt{\lambda(m_{K^0}^2,m_{\pi^0}^2,m_S^2)}}{64\pi m_{K^0}^3}\(\frac{m_{K^0}^2-m_{\pi^0}^2}{m_s-m_d}\)^2 \tau_{K_S},
\eea
where the kinematic function $\lambda(x,y,z) = x^2+y^2+z^2-2(xy+yz+zx)$ and $f_0(q^2)$, $A_0(q^2)$ are the relevant scalar $B$ meson form factors \cite{Ball:2004ye, Straub:2015ica} while $f_{0,K}(q^2)$ is the kaon form factor \cite{Marciano:1996wy}.


The contributions to the branching ratio of the different $B_s$ decays in the dark scalar model are given by~\cite{Datta:2019bzu, Felkl:2023ayn}
\beq 
\mathcal{B}(B_s \to \gamma \gamma) = \frac{|g_{bs}|^2 \kappa^2}{64 \pi} \frac{f_{B_s}^2 m_{B_s}^7}{m_b^2(m_{B_s}^2-m_S^2)^2}\tau_{B_s}\,,
\eeq 
\beq 
\mathcal{B}(B_s \to \mu^+ \mu^-) = \frac{|g_{bs}|^2 \eta_d^2 m_\mu^2}{32 \pi v^2} \frac{f_{B_s}^2 m_{B_s}^5}{m_b^2(m_{B_s}^2-m_S^2)^2}\(1-\frac{4m_{\mu}^2}{m_{B_s}^2}\)^{3/2}\tau_{B_s}\,,
\eeq 
\beq 
\mathcal{B}(B_s \to \nu\bar\nu) = \frac{|g_{bs}|^2 g_D^2 (1-|U_{\mu 4}|^2)^2}{128 \pi} \frac{f_{B_s}^2 m_{B_s}^5}{m_b^2(m_{B_s}^2-m_S^2)^2}\(1-\frac{4m_{\nu 4}^2}{m_{B_s}^2}\)^{3/2}\tau_{B_s}\,,
\eeq 
where $f_{B_s}$ is the $B_s$ meson decay constant \cite{FlavourLatticeAveragingGroupFLAG:2021npn}. 
\section{Meson mixing}
The scalar contribution to $B_s^0$ mixing is given by \cite{Datta:2019bzu}
\label{sec:mixing}
\beq 
\Delta M_{B_s} =-\frac{5}{12}\frac{|g_{bs}|^2}{m_{B_s}^2-m_S^2}f_{B_s}^2 m_{B_s}\,.
\eeq
A similar expression applies for kaon mixing.

\section{Coherent scattering}
\label{sec:coherent}
The interaction of the dark scalar to a nucleus with atomic number $Z$ and mass number $A$ is given by
\beq 
\mathcal{L}_{SN} = C_N \bar\psi_N \psi_N S\,,
\eeq 
where $\psi_N$ is the spinor of the nucleus and the effective coupling $C_N$ in terms of the scalar coupling to protons ($C_p$) and neutrons ($C_n$) reads
\beq
C_N = Z C_p + (A-Z) C_n\,.
\eeq 
The proton and neutron couplings are related to the quark-scalar couplings by
\beq 
C_p = \frac{m_p}{v}\(\eta_c f_c^p + \eta_t f_t^p + \sum_d \eta_d f_d^p\),~~ C_n = \frac{m_n}{v}\(\eta_c f_c^n + \eta_t f_t^n + \sum_d \eta_d f_d^n\)\,,
\eeq 
where $m_p = 938.3~\rm{MeV}$, $m_n = 939.6~\rm{MeV}$ are the proton and neutron masses and $f^{p,n}$ are the respective form factors~\cite{Crivellin:2013ipa, Hoferichter:2015dsa, Junnarkar:2013ac}. 

The differential cross section of the coherent scattering process, $\nu_\mu + N \to \nu_4 + N$, is given by
\beq 
\frac{d\sigma_S}{dT} = \frac{g_D^2}{16\pi}|U_{\mu 4}C_N|^2 (1-|U_{\mu 4}|^2)^2 \frac{(2M + T)(m_{\nu 4}^2 + 2MT)}{E_{\nu_\mu}^2 (m_S^2 + 2MT)^2} F^2(T)\,,
\eeq 
where $E_{\nu_\mu}$ is the muon neutrino energy, $M$ is the mass of the nucleus, $T$ is the recoil energy, and $F(T)$ is the nuclear form factor~\cite{Helm:1956zz}.

\section{$(g-2)_\ell$}
\label{sec:g2}
The dominant scalar contribution to $(g-2)_\ell$ via the Barr-Zee diagram is given by~\cite{Davoudiasl:2018fbb}
\begin{equation}
\delta (g-2)_{\ell}^{S\gamma\gamma} \approx \frac{\eta_d}{4 \pi^2} \frac{\kappa m_\ell^2}{v} 
\ln  \frac{\Lambda}{m_S} \,, 
\label{g-2}
\end{equation}
where $\Lambda$ is the cutoff scale of the theory that generates the effective $S\gamma\gamma$ coupling. 

The subdominant contribution from the one-loop correction
is given by~\cite{Leveille:1977rc}
\begin{equation}
\delta (g-2)_{\ell}^{(\text{1-loop)}}=\frac{\eta_d^2}{8\pi^2}{m_\ell^2 \over v^2}\int_0^1 dz\frac{(1+z)(1-z)^2}{(1-z)^2+ z/r_\ell^{2}} \,,
\end{equation}
where $r_\ell=m_\ell/m_S$.

\end{document}